\documentclass[useAMS,usenatbib,usegraphicx]{mn2e}

\newcommand{\vx}{\mbox{\bf {x}}}
\newcommand{\vy}{\mbox{\bf {y}}}
\newcommand{\vk}{\mbox{\bf {k}}}
\newcommand{\vq}{\mbox{\bf {q}}}
\input epsf

\title{Cross Terms and Weak Frequency Dependent Signals in the CMB Sky}

\author[C. Hern\'andez--Monteagudo \& R.A.Sunyaev]{C.
Hern\'andez--Monteagudo$^{1}$\thanks{E-mail:chm@MPA-Garching.MPG.DE} and
R.A.Sunyaev$^{1,2}$\\
$^{1}$Max-Planck Institut f\"ur Astrophysik, Karl-Schwarzschild-Str. 1,
D-85740 Garching, Germany\\
$^{2}$Space Research Institute (IKI), Profsoyuznaya 84/32, Moscow 117810, Russia \\
}
\begin{document}

\date{}

\pagerange{\pageref{firstpage}--\pageref{lastpage}} \pubyear{2003}

\maketitle

\label{firstpage}

\begin{abstract}

In this paper, we study the amplification of weak frequency dependent
signals in the CMB sky due to their cross correlation to intrinsic
anisotropies. In particular, we center our
attention on mechanisms generating some weak signal, of peculiar
spectral behaviour, such as 
resonant scattering in ionic, atomic or molecular lines,
thermal SZ effect or extragalactic foreground emissions, whose typical 
amplitude (denoted by $\epsilon$) is sufficiently smaller than the 
intrinsic CMB fluctuations.
 We find that all these effects involve either the autocorrelation of 
anisotropies generated during recombination {\bf ($z_{rec}$)}
or the cross-correlation of
those anisotropies with fluctuations arising at redshift $z_i$. The
former case accounts for the slight blurring of original anisotropies
generated in the last scattering surface, and shows up in the small
angular scale (high multipole) range.
The latter term describes, instead, the generation of new
anisotropies, and is non-zero only if fluctuations generated at redshifts 
$z_{rec},\;z_i$ are correlated. The degree of this correlation can be computed
under the assumption that density fluctuations were generated as standard 
inflationary models dictate and that they evolved in time according to
linear theory. In case that the weak signal is frequency dependent, (i.e.,
the spectral dependence of the secondary anisotropies is distinct from
that of the CMB),
we show that, by substracting power spectra at different frequencies, 
it is possible to
avoid the limit associated to Cosmic Variance and unveil weaker terms linear
in $\epsilon$. We find that the correlation term shows a different spectral
dependence than the squared ($\propto \epsilon^2$) term considered
usually, making its extraction particularly straightforward for the thermal SZ
effect. Furthermore,
we find that in most cases the correlation terms are particularly relevant
at low multipoles due to the ISW effect
 and must be taken into account when characterising the
power spectrum associated to weak signals in the large angular scales.

\end{abstract}

\begin{keywords}
cosmic microwave background --  large scale structure of
Universe -- galaxies: clusters: general -- methods: statistical

\end{keywords}

\section{Introduction}

Standard theories state that the field of density perturbations arising
after the inflationary epoch, 
($\delta(\vx) \equiv (\rho(\vx)-\bar{\rho})/\bar{\rho}$, with $\bar{\rho}$ the
average density), should be
gaussian,
homogeneous and isotropic, \citep{Guth81, Staro81, Mukh82, Linde83}.
The Fourier modes of this field ($\delta_{\vk}$) 
are predicted to have independent real
and imaginary components, which should be gaussian distributed 
from a scale-invariant power spectrum, 
(Harrison--Zel'dovich, (HS), \citep{HS}), 
i.e.,
$\langle \delta_{\vk} \delta_{\vq}^* \rangle = (2\pi)^3 P(k) 
\delta_D (\vk+\vq)$, with $P(k) \propto k$. This
power spectrum determines the properties of the spatial correlation of the
perturbation field, since it is the mere Fourier transform of the correlation
function.
These perturbations are {\em small} compared to the homogeneous background,
($|\delta| \ll 1$), but grow up due to gravitational instabilities. This
growth is independent for each mode, i.e., mode coupling can be neglected,
as long as the perturbations remain small and linear theory can be applied.

From the observational point of view, the first test ground 
for this perturbation field is the study of the temperature anisotropies
of the Cosmic Microwave Background (CMB).
Most of these temperature fluctuations
were generated by energy density inhomogeneities 
in the universe during the epoch at which most electrons and protons
 recombined to
form hydrogen and radiation decoupled from matter, (last scattering surface,
LSS). At this stage, the density inhomogeneities were still under linear 
regime, provided that the amplitude of
typical measured CMB temperature fluctuations are one part in 
one hundred thousand, (e.g. \citet{smoot91,Bennett03}). In their transit from the 
LSS towards us, the
CMB photons witnessed the matter collapse and formation of 
non linear structures such as galaxies, clusters of galaxies, filaments
and superclusters of galaxies that today conform the visible universe.
The crossing through these scenarios imprinted on the CMB photons
new temperature anisotropies, which are usually labelled as {\it secondary}.
The amplitude of these secondary anisotropies is, in many cases,
few orders of magnitude below the level of the primary ones generated
in the LSS. However, the different spectral behaviour 
of some of them might help in the distinction from the primary.
 In this context, the presence of foregrounds, galactic
or extragalactic, with their own spectral dependence, will make the picture
furtherly more complicated.

Consequently, a major issue in current CMB science is the accurate component
separation in future microwave maps. From the observation point of view, 
a set of space and groundbased 
experiments  with unprecendented sensitivity and angular resolution, 
{\em counting with several broad band detectors spread in appropiate frequency 
ranges}, are being proposed or already under development, 
(e.g., 
Planck\footnote{Planck's URL site: \\
{\it http://www.rssd.esa.int/index.php?project=PLANCK}}, 
ACT\citep{ACT04}, South Pole Telescope\footnote{South Pole Telescope's URL
site:\\
{\it http://astro.uchicago.edu/spt/}},QUIET{\footnote{Key URL site for
QUITE:\\
{\it http://cfcp.uchicago.edu/capmap/QUIET.htm}} or
CMBPOL\footnote{CMBPOL's URL site: \\
 {\it http://www.mssl.ucl.ac.uk/www\_astro/submm/CMBpol1.html}}).
In the theoretical side, 
new analysis techniques of temperature maps, based both in real 
and Fourier space, and dealing with second (correlation function, 
power spectrum) and higher order momenta of quantities derived from
the temperature field are being developed 
and tested on simulated data. Nevertheless,
the two main limiting factors in this task will be {\it i)} the 
instrumental noise and instrument systematics 
and {\it ii)} the cosmic variance,
associated to the fact that our characterization of the universe is
statistical, but based on a single realization of it.

In this work, we study the spatial correlation of density fluctuations
in the universe, and how this reflects in the CMB angular power spectrum.
These aspects must be taken into account if an accurate characterization of 
the CMB power spectrum is to be achieved, particularly at the large angular 
(low multipole) scales. This study also allows us to propose a method
that uses observations in different frequencies and combines power 
spectra in such a way that {\em avoids the limitation imposed by
cosmic variance}, and unveils weak signals whose amplitude $\sigma_w$ is in
the range $\sigma_N >\sigma_w >\sigma_N \left( \sigma_N / \sigma_t \right)$,
where $\sigma_N$ is the experimental noise amplitude and $\sigma_t$ is the
typical amplitude of a dominant signal $t$ which is assumed to be totally
correlated to the weak signal $w$. 
This approach was already utilized in \citet{Basu} when 
charactering the effect of metal atoms and ions on the CMB during the
secondary ionization.\\

In Section 2 we outline our method, which we apply in Section 3
on particular physical mechanisms generating secondary fluctuations
in two different cosmological scenarios. In Section 4
 we comment our results and conclude.\\

\section{Comparing Second Order Momenta}

\subsection{The Flat Case}

Our starting point will be the superposition of two signals, $g_1(\nu)\;t_1$ 
and $g_2(\nu)\;t_2$, whose amplitudes will show, {\it a priori}, 
a different frequency ($\nu$) dependence.
 In real space they give rise to
\begin{equation}
T(\vx,\nu) = g_1(\nu) t_1(\vx) + g_2(\nu) \epsilon t_2(\vx) + N_{\nu}(\vx),
\label{eq:temp1}
\end{equation}
and analogously, in Fourier space, to
\begin{equation}
T_{\vk}(\nu) = g_1(\nu) t_{1,\vk} + g_2(\nu)\epsilon t_{2,\vk} + N_{\nu,\vk},
\label{eq:temp1b}
\end{equation}
where $\vk$ is the Fourier mode under consideration. If we assume that 
$g_1(\nu_1)t_1$ and $g_2(\nu_2)t_2$ are of similar amplitude, 
then the parameter $\epsilon$ gives the relative amplitude of both signals
, and for the cases considered below, we shall take 
$\epsilon \ll 1$. $N_{\nu}$ is the noise component present in the map.
Now let us assume that the experiment is able to observe at two different
frequencies $\nu_1,\nu_2$. Defining $f\equiv g_1(\nu_1) / g_1(\nu_2)$, we find
that:
\[
\delta \left( T^2 (\vx,\vy) \right) \equiv \; T(\vx,\nu_1) T(\vy,\nu_1)-
			f^2T(\vx,\nu_2)T(\vy,\nu_2) \; = 
\]
\[
\phantom{xx}	\epsilon\;\;
	 g_1(\nu_1)g_2(\nu_2 ) 
				\left[ 1 - \frac{g_1(\nu_1)}{g_1(\nu_2)}
						\frac{g_2(\nu_2)}{g_2(\nu_1)} \right]
  	\left[t_1(\vx) t_2(\vy) + t_1(\vy) t_2(\vx) \right]
\]
\[ 
\phantom{xx} 
 + \epsilon^2\;\; g_2^2(\nu_1)\left[1-\left(\frac{g_1(\nu_1)}{g_1(\nu_2)}
						\frac{g_2(\nu_2)}{g_2(\nu_1)} \right)^2 \right]
					t_2(\vx)t_2(\vy)
\]
\[
\phantom{xx} 
+ N_{\nu_1}(\vx)N_{\nu_1}(\vy) - f^2\;N_{\nu_2}(\vx)N_{\nu_2}(\vy)
\]
\begin{equation}
\phantom{xx} 
 + {\cal O} \left[N_{\nu_1},N_{\nu_2} \right],
\label{eq:dTsq1}
\end{equation}
\vspace{.4cm}

\noindent or, in Fourier space,
\\

\[
\delta \left( T_{\vk,\vq}^2 \right) \equiv T_{\vk}(\nu_1)T_{\vq}(\nu_1)-
							f^2T_{\vk}(\nu_2)T_{\vq}(\nu_2)=
\]
\[
\phantom{xx}
	\epsilon\; g_1(\nu_1)g_2(\nu_1)\left[1-\frac{g_1(\nu_1)}{g_2(\nu_2)}
								\frac{g_2(\nu_2}{g_2(\nu_1)}\right]
	\left[ t_{1,\vk}t_{2,\vq} + t_{2,\vk}t_{1,\vq} \right] 
\]
\[
\phantom{xx}
+	\epsilon^2\;\; g_2^2(\nu_1)\left[1-\left(\frac{g_1(\nu_1)}{g_2(\nu_2)}
								\frac{g_2(\nu_2}{g_2(\nu_1)}\right)^2\right]
			\;\;t_{2,\vk}t_{2,\vq} 
\]
\[
\phantom{xx}
+  N_{\nu_1,\vk}N_{\nu_1,\vq} - f^2\;N_{\nu_2,\vk}N_{\nu_2,\vq} 
\]
\begin{equation}
\phantom{xx}
+  {\cal O}\left[ N_{\nu_1}, N_{\nu_2} \right].
\label{eq:dTsq2}
\end{equation}
${\cal O}\left[ N_{\nu_1},N_{\nu_2} \right]$ in both equations refers
 to cross terms of the
noise field with all the other components at a given frequency.
These two equations should be compared to the squared difference map,
($\left( \delta T \left(\vx,\vy\right) \right)^2 $, 
$\left( \delta T_{\vk,\vq} \right)^2$), given by:
\[ 
\left( \delta T \left(\vx,\vy\right) \right)^2 \equiv 
\]
\[
\phantom{xx}
\left( T(\vx,\nu_1)-fT(\vy,\nu_2) \right)
					\left(T(\vx,\nu_1)-fT(\vy,\nu_2) \right) =
\]
\[
\phantom{xx}
\epsilon^2 \; \large( \;\left[ g_2(\nu_1)\;t_2(\vx)\right]^2 \;  +
	\left[f\;g_2(\nu_2)\;t_2(\vy) \right]^2 - 
\]
\[
\phantom{xxxxxxxxxxxx}
			2\;f\; g_2(\nu_1)g_2(\nu_2)t_2(\vx)t_2(\vy)\;\large)  \; +
\]
\begin{equation}
\phantom{xx}
N_{\nu_1}^2(\vx) + N_{\nu_1}^2(\vy) + 
			{\cal O}\left[ N_{\nu_1},N_{\nu_2}\right],
\label{eq:sqdT1}
\end{equation}
in real space, and 
\[ 
\left( \delta T_{\vk,\vq} \right)^2 \equiv 
\]
\[
\phantom{xx}
\left( T_{\vk}(\nu_1)-fT_{\vq}(\nu_2) \right)
					\left(T_{\vk}(\nu_1)-fT_{\vq}(\nu_2) \right) =
\]
\[
\phantom{xx}
\epsilon^2 \;\; \large( \left[ t_{2,\vx}g_2(\nu_1)\right]^2 +
	\left[f t_{2,\vy} g_2(\nu_2)\right]^2 - 
\]
\[
\phantom{xxxxxxxxxxxx}  
			2\;f\;g_2(\nu_1)g_2(\nu_2)t_{2,\vx}t_{2,\vy}\;\large) \; +
\]
\begin{equation}
\phantom{xx}
N_{\nu_1,\vk}^2 + N_{\nu_1,\vq}^2 + 
			{\cal O}\left[ N_{\nu_1},N_{\nu_2}\right]
\label{eq:sqdT2}
\end{equation}
in Fourier space.

It is clear that for $\epsilon \ll 1$, 
$\delta \left( T^2  (\vx,\vy)\right) $  or 
$\delta \left( T_{\vk,\vq}^2 \right)$
 are much more sensitive to the weak signal $\epsilon t_2$ than 
$\left( \delta T \left(\vx,\vy\right) \right)^2 $ or
 $\left( \delta T_{\vk,\vq} \right)^2 $.
The obvious difference is the term linear in $\epsilon$ present in 
eqs.(\ref{eq:dTsq1},\ref{eq:dTsq2}). However, in the
context of Cosmology and CMB, one counts with only one single realization
of the Universe, and the quantities defined above
as $\delta \left( T^2 (\vx,\vy)\right) $ or 
$\delta \left( T_{\vk,\vq}^2 \right)$ must be averaged either in real
or Fourier space, in order to acquire some statistical meaning, (i.e., if
averaged under certain conditions, they yield estimates of the
correlation function and the power spectrum, respectively). 
After this average, the term
linear in $\epsilon$ becomes proportional to $\langle t_1 \; t_2\rangle$,
and will not average out {\em if and only if both signals $t_1$, $t_2$ are 
correlated}, at least to some extent. Therefore, in order for this cross term
to be of any utility, {\em both the dominant and the weak signals must be 
correlated}. We shall show 
below that this is indeed the case for signals coupled
to linear fluctuations of the density field generated after inflation.
Another point to remark is that, because of substracting
quantities computed from the same maps, one {\em exactly} cancels the
dominant signal, {\em leaving no room for the uncertainty due to the 
cosmic variance associated to it.} This allows the weak signal be
under the limit imposed by the cosmic variance of the dominant one. \\

As mentioned in the Introduction, in linear theory all
Fourier modes $\delta_{\vk}$ of the density fluctuations evolve
independently according to a growth factor $D(\eta )$ ($\eta $ is
conformal time) which is dependent on the cosmological parameters
of our universe. These modes are all independent, and for
reasons associated to the homogeneity and isotropy,
must depend exclusively on the modulus of the $\vk$ vectors, $k$. This
allows writing the power spectrum as
$\langle \delta_{\vk} \delta_{\vq} \rangle = (2\pi)^3 P(k) \delta_D(\vk+\vq)$.
In an analogous way, the averages of the product of all pair of
quantities depending linearly
on $\delta_{\vk}$ will be proportional to the power spectrum. 
This applies practically to 
all perturbations of physical quantities, such as peculiar velocities
or gravitational potentials, that are responsible for the generation of
temperature anisotropies in the CMB. \\

The average in our maps will be performed in the real space in such a way that
the distance between $\vx$ and $\vy$ is kept constant. In Fourier
space, we shall take\footnote{Note that, for real
signals, $\delta_{-\vk} = \delta^*_{\vk}$.} $\vq$ equal to $-\vk$,
 fix the modulus (k) and average over the
mode phases. The former will yield the correlation function, the
latter the power spectrum.
This average also removes all cross terms in noise. Furthermore, 
if we assume that the statistical properties of
noise have been characterized, then it is possible to substract the
{\em expectations} for the terms quadratic in noise
 in eqs.(\ref{eq:dTsq1},\ref{eq:dTsq2}),
and the residuals of this substraction can be treated as random variables.
These random residuals should be regarded as the {\it effective} noise in
our correlation function or power spectrum estimates, and will be denoted
by $\Delta N_{\nu,\vx-\vy}$ and $\Delta N_{\nu,k}$:
\begin{eqnarray}
\Delta N_{\nu,\vx-\vy} &  \equiv & {\cal E} \bigl( \langle N_{\nu}(\vx) N_{\nu}(\vy) \rangle \bigr)
		- \langle N_{\nu}(\vx) N_{\nu}(\vy) \rangle_{EXP} 
\label{eq:quaderr1} \\
\Delta N_{\nu,k} &  \equiv & {\cal E} \bigl( \langle N_{\nu,\vk}  N_{\nu,-\vk} \rangle \bigr)
		- \langle N_{\nu,\vk} N_{\nu,-\vk} \rangle_{EXP}, 
\label{eq:quaderr1b}
\end{eqnarray}
where ${\cal E}$ and the label {\it EXP} denote {\it estimated on the map} and
{\it expected} values, respectively. We are assuming
that noise in different frequencies is uncorrelated. For the
case of gaussian white noise, it is easy to prove that 
$\langle \Delta N_{\nu,\vx-\vy} \rangle = 
			\langle \Delta N_{\nu,k} \rangle = 0$,
(assuming a correct characterization of noise), and that
\begin{eqnarray}
\langle \Delta N_{\nu,\vx-\vy}^2 \rangle & = & \frac{2}{n}
		\langle N_{\nu}(\vx) N_{\nu}(\vy) \rangle_{EXP}^2
\label{eq:quaderr2} \\
 \langle \Delta N_{\nu,k}^2 \rangle & = & \frac{2}{n}
		\langle N_{\nu}(\vk) N_{\nu}(-\vk) \rangle_{EXP}^2.
\label{eq:quaderr2b}
\end{eqnarray}
$n$ is the number of points, either in real of Fourier space, used when
estimating the averages\footnote{For {\em white} gaussian noise, 
if $\vx \neq \vy$,
then $\langle \Delta N_{\nu,\vx-\vy}^2 \rangle = \langle
(N_{\nu}(\vx))^2\rangle \langle (N_{\nu}(\vy))^2 \rangle/ n$ }. 
Having this in mind, we can perform the averages and 
rewrite eqs.(\ref{eq:dTsq1},\ref{eq:dTsq2}) like
\[
{\cal E} \bigl( \langle \delta \left( T^2 (\vx,\vy) \right) \rangle \bigr) =
\]
\[
 \epsilon \;\;
	 g_1(\nu_1)g_2(\nu_2 ) 
				\left[ 1 - \frac{g_1(\nu_1)}{g_1(\nu_2)}
						\frac{g_2(\nu_2)}{g_2(\nu_1)} \right]
  	2\;\langle t_1(\vx) t_2(\vy) \rangle
\]
\begin{equation}
\phantom{xxxxxxxxxx}
\pm \Delta_N \; + \;{\cal O}\left[ \epsilon^2 \right]
\label{eq:dTsq3}
\end{equation}
and
\[
{\cal E} \bigl( \langle \delta \left(T_{\vk,-\vk}^2 \right)  \rangle \bigr)\; =
\]
\[
	\epsilon\; g_1(\nu_1)g_2(\nu_1)\left[1-\frac{g_1(\nu_1)}{g_1(\nu_2)}
								\frac{g_2(\nu_2)}{g_2(\nu_1)}\right]
	\langle t_{1,\vk}t_{2,-\vk} + t_{2,\vk}t_{1,-\vk} \rangle \; 
\]
\begin{equation}
\phantom{xxxxxxxxxx}
\pm \Delta_N \; + \; {\cal O}\left[ \epsilon^2 \right].
\label{eq:dTsq4}
\end{equation}
$\Delta _N$ is the residual noise contribution, 
$\Delta_N^2=\langle \Delta N_{\nu_1}^2 \rangle + 
					f^2\langle \Delta N_{\nu_2}^2 \rangle$ in both
real and Fourier space, (eqs.(\ref{eq:quaderr2}--\ref{eq:quaderr2b})).
From this equation, one can see that the approach proposed here will
be sensitive to $\epsilon \;t_2$ if:
\begin{equation}
\omega\;\epsilon > 
					\frac{\sigma_N^2}{\sigma_t^2} \times
		\frac{\sqrt{\frac{2}{n}\left( 1+f^2\right)}} 
		{2\cdot\left( 1-\frac{g_1(\nu_1)g_2(\nu_2)}{g_1(\nu_2)g_2(\nu_1)}\right) },
\label{eq:rangeep}
\end{equation}
with $\sigma_N^2 \equiv \langle N_{\nu}^2(\vx) \rangle$ taken equal for the two
frequencies and $\sigma_t^2 \equiv g_1^2(\nu_1)
\langle t_1^2 \rangle \sim g_2^2 (\nu_2)\langle t_2^2 \rangle $.
The factor $\omega$ accounts for the cross correlation between $t_1$ and
$t_2$, i.e., $\omega \equiv \langle t_1 \cdot t_2 \rangle / \sigma_t^2$,
(note that in the absence of correlation, $\omega = 0$ since we are taking
$\langle t_1 \rangle = 0$ by construction).
Let us remark that the limit on $\omega\epsilon$ is roughly an
order in $\sigma_N / \sigma_t$ beyond the limit imposed on $\epsilon$ by eqs.
(\ref{eq:sqdT1},\ref{eq:sqdT2}). Note that in the case of
similar frequency dependence for the two signals, ($g_1(\nu) \simeq g_2(\nu)$),
this method cannot work. For similar reasons, if $g_1(\nu ) \neq g_2(\nu )$, 
then it should be possible, {\it a priori}, to perform
 as many consistency checks in different frequencies as the
instrument permits, since the correlation term should vary its amplitude
as dictated by the frequency dependent term
\begin{equation}
\Gamma (\nu_1, \nu_2) \equiv 
	g_1(\nu_1)g_2(\nu_1)\left[1-\frac{g_1(\nu_1)}{g_1(\nu_2)}
	\frac{g_2(\nu_2)}{g_2(\nu_1)}\right].
\label{eq:freqterm}
\end{equation}

From this formalism, it follows that the importance of this approach
relies {\it i)}  on the amplitude of the cross-correlation between the
signals under consideration and {\it ii)} on their spectral dependence.
Let us remark as well that this method is sensitive to the relative
sign of the two signals.
In the context of the CMB, this correlation will preferrably show up in the
low multipole range: at these large angular scales the 
instrumental sensitivity performs best, but the removal of galactic foregrounds
becomes particularly difficult. In the next section,
we shall address several scenarios where this correlation may be relevant,
and discuss under which conditions the method proposed here becomes
useful.\\

The approach outlined here is complementary, but different, to that used
in, e.g., \citet{banday96}, \cite{kneissl97}, \citet{jal00}, 
and more recently, \citet{BC03}, \citet{pablo03} and \citet{scjal04}.
In all those cases, the weak signal ($\epsilon t_2$) was not considered
to be correlated to the dominant signal, but it was  
cross-correlated to 
an external template: this cross-correlation retained only the term
linear in $\epsilon$, and hence no substraction was required.\\

Hereafter, the term proportional to $\epsilon$ will be referred to
as the {\it linear} or {\it cross} term, whereas the term
proportional to $\epsilon^2$ will be denoted as the {\it squared} term.\\

\subsection{Correlations Projected on the Sphere}

In this subsection we briefly outline the formalism that describes the 
analysis of temperature fluctuations in the CMB. It is customary
to work in the spherical  Fourier space, 
in which the coefficients $a_{l,m}$'s define
a temperature field in the celestial sphere through the following
decomposition on spherical harmonics:
\begin{equation}
\frac{\delta T}{T_0} (\theta, \phi) = 
	\sum_{l,m} a_{l,m} \; Y_{l,m} (\theta, \phi).
\label{eq:alm_des}
\end{equation}

The power spectrum for an arbitrary temperature field is obtained 
after averaging the Fourier coefficients,
\begin{equation}
C_l \equiv \langle a_{l,m} a_{l,m}^* \rangle.
\label{eq:cldef}
\end{equation}

Having this in mind, the 
analysis of weak signals outlined in the previous section translates
into the spherical case as
\begin{equation}
 \delta C_l = 2 \epsilon \Gamma(\nu_1,\nu_2)\langle a_{l,m} (a_{l,m}^{weak})^*
\rangle + {\cal O} [\epsilon^2].
\label{eq:deltaC_l}
\end{equation}

However, when computing this correlations, it will be convenient to
express the $a_{l,m}$'s as integrals in the flat Fourier space. Indeed,
the temperature field can be decomposed in Fourier modes as 
(e.g.,\citet{HuS95}):
\[
\Delta\left(\vk,{\bf n},\eta_0 \right) = 
		\int d\vx \frac{\delta T}{T_0}(\vx,{\bf n},\eta_0 )
			 \; e^{-i\vk\vx} =
\]
\begin{equation}
\phantom{xxxxxxxxxx}
	\sum_l (-i)^l \left( 2l+1 \right) P_l (\mu ) \Delta_l (\vk, \eta_0 ),
\label{eq:Leg1}
\end{equation}
with $\mu=\hat{ {\bf k}}\cdot \hat{ {\bf n}}$, and $\hat{{\bf n}}$ is
the pointing vector on the sky given by $(\theta,\; \phi)$. $\eta_0$ denotes
the conformal time evaluated at the present epoch.
The last step shows the
expansion on a Legendre polynomial basis, and assumes implicitely that
perturbations are {\em axially symmetric} about $\vk$, (e.g., \citet{MB95}).
 From this, it is straightforward
to show that, for $\vx = 0$, the $a_{l,m}$ multipoles can be written as:
\begin{equation}
a_{l,m} = \left( -i \right)^l \; 4\pi\; 
	\int d\vk Y_{l,m}^* (\hat{k})\; \Delta_l (\vk, \eta ).
\label{eq:alm}
\end{equation}

In linear theory, $\Delta_l (\vk, \eta ) = \Delta_l (k, \eta )
					\psi_i \left( \vk \right)$, with 
$\psi_i \left( \vk \right)$  the initial scalar perturbations and
$\langle \psi_i \left( \vk \right) \psi_i \left( \vq \right) \rangle =
P_{\psi}\left( k \right) \left( 2\pi \right)^3 \delta_D
\left( \vk + \vq \right)$ the initial scalar perturbation power spectrum.
It turns out that, after integrating the Boltzmann equation,
the mode $\Delta\left(\vk,{\bf n},\eta_0 \right)$ can often be written as
a line-of-sight (LOS) integral of some sources dependent on 
$\vk$ and $\eta$, $S(\vk,\eta)$, \citep{SZald96}:
\begin{equation}
\Delta\left(\vk,{\bf n},\eta_0 \right) = \int d\eta\; e^{i\;k\mu[\eta_0-\eta]}
	S(\vk,\eta),
\label{eq:source1}
\end{equation}
where the sources can be related to the velocity, potential and/or density
perturbation modes.
After using the Rayleigh expansion for the exponential in 
equation (\ref{eq:source1}),
it is easy to show that the multipoles $\Delta_l (\vk, \eta_0 )$ can
be expressed as \citep{SZald96}:
\begin{equation}
\Delta_l (\vk, \eta_0 ) \propto \int d\eta j_l[k(\eta_0-\eta)] S(k,\eta).
\label{eq:delta_l_2}
\end{equation}
This is only correct if the source term has no $\mu$ dependence,
$S(\vk, \eta) = S(k, \eta)$. Otherwise the integral along the LOS is 
projected on spherical Bessel functions of different order,
(i.e. $\Delta_l$ is an integral of $j_{l+1}$ and $j_{l-1}$ if $S(\vk, \eta)
\propto \mu$). 
In all cases
considered here, the sources will be $\mu$ independent, and 
eq. (\ref{eq:delta_l_2}) will be used. 

This expresion of $\Delta_l (\vk, \eta_0 )$ also allows us to make some
predictions regarding the multipole range where the cross-correlation term
$\langle a_{l,m} (a_{l,m}^{weak})^*\rangle$ will be relevant. 
The formal way to see this is through the integral defining $\delta C_l$:
\[
\delta C_l \propto \int dk \; k^2\; P_{\psi} (k) \;S_1(k,\eta_1)
 \; S_2(k, \eta_1) \;  \times
\]
\begin{equation}
\phantom{xxxxxxxxxxxxxxxxxx}
j_l (k[\eta_0 - \eta_1]) j_l(k[\eta_0 - \eta_2]).
\label{eq:delta_C_l_2}
\end{equation}
In this equation, we have assumed that the two signals have been generated
at conformal times $\eta_1$ and $\eta_2$ (with $\eta_1 > \eta_2$). For 
a fixed $l$, we have that $j_l(x) \sim 1$ if $x \sim l$. For $x \ll 1$,
$j_l(x) \sim x^l$, and $j_l(x) \sim \cos (x -l\pi/2 -\pi/4) / x$ if $x \gg l$.
From this it is easy to see that, for a fixed $l$,
the spherical Bessel functions will be close to unity if $k \sim k_1 \equiv l/\delta \eta_1$, $k \sim k_2 \equiv l/\delta \eta_2$ in each case, 
($\delta \eta_i \equiv \eta_0 - \eta_i$, $i=1,2$). In practice, this means
that, given that $k_1 > k_2$, for the $k$ range for which $j_l(k \delta \eta_2)$ is unity $k \delta \eta_1 < l$, so
that, for the very low $k$'s (and hence very
low $l$'s), $j_l(k \delta \eta_1)$ will approach to zero if 
$ k\delta \eta_1 \ll 1$. This reflects the fact that such modes {\it do not enter} in the angular scales given by $l$, and it is easy to show that this will
take place predominantly in multipoles below 
$l_{min} \equiv (\eta_0 - \eta_2) / (\eta_0 - \eta_1) $.
On the other hand, for the
$k$ range for which $j_l(k \delta \eta_1) \sim 1$, we then have that
$j_l (k \delta \eta_2) \sim \cos (k\delta\eta_2 - l\pi/2 - \pi/4) / 
(k\delta\eta_2)$ if $k\delta \eta_2 \gg l$. Hence, the phase difference 
between both Bessel functions will become important if $k (\eta_1 - \eta_2) \sim 2\pi$, or equivalently, for $l_{max} \sim 2\pi (\eta_0 - \eta_2) / (\eta_1 - \eta_2) $. 
$l_{max}$ stands for the multipole at which we expect a change in
the cross-correlation structure between two relatively nearby signals.
However, we may find scenarios in which both signals are
so distant that $l_{max} \sim 1$, and for which this analysis cannot be 
applied. Also, we must keep in mind the caveat that we are ignoring
the $k$ dependence of the sources, which condition the actual amplitude
of the correlation.\\

\subsection{Frequency Dependence of the Cross Terms}

We next focus on the frequency dependence of the $\delta C_l$'s.
This method is based upon the assumption that dominant and weak signals
have different spectral dependence. This translates into a frequency
dependence of the $\delta C_l$'s given by:

\[
\delta C_l = \left(g_2(\nu_1) - g_2(\nu_2) \right)\; \times
\]
\begin{equation}
\phantom{x} \biggl( 
  2\;\epsilon \langle a_{l,m}\; (a_{l,m}^{weak})^*\rangle \; + \;\epsilon^2
\bigl( g_2(\nu_1) + g_2(\nu_2) \bigr)
         \langle | a_{l,m}^{weak} |^2 \rangle \biggr),
\label{eq:deltaC_l_nudp}
\end{equation}
where we have taken $t_1$ to be the primordial CMB fluctuations
and hence $g_1(\nu) = 1 = const$. 
This equation 
shows the frequency dependence of the $\delta C_l$'s and 
also manifests
{\em the different behaviour of the correlation term and the squared
term with respect to $\nu$}. That is, if we define $\Delta (g) \equiv
g_2(\nu_1) - g_2(\nu_2)$ and $\Delta (g^2) \equiv g_2^2(\nu_1) -
g_2^2(\nu_2)$, then the (linear) cross term is proportional to $\Delta (g)$,
whereas the squared term is multiplied by  $\Delta (g^2)$, e.g., the latter
is more sensitive to big changes in $g(\nu)$. This different behaviour
should  motivate the choice of observing frequencies in order to distinguish
the contribution of both terms.
\\

\subsection{Relative Sign Dependence of Weak and Dominant Signals}

Since the cross term couples different signals, it is sensitive to the
relative sign or phase present between them. That is, it is sensitive to
whether both signals are correlated or anticorrelated. This sign depends
upon the physical processes relating both signals and their particular
spectral dependence, and can be different in different $l$ ranges.\\
    
In the case of
 the thermal Sunyaev-Zel'dovich effect (hereafter tSZ,
\citet{sunyaev80}), we shall find that, for the low frequencies
for which the effect decreases the CMB brightness, ($\nu < 218$ GHz),
the tSZ will be anticorrelated to the intrinsic CMB temperature fluctuations
(caused mainly through the late ISW effect), whereas for $\nu > 218$ GHz
both signals become correlated.\\

For resonant scattering, at high $l$'s, we shall see that blurring
of original CMB anisotropies dominates ($\delta C_l < 0$), whereas at low
multipoles generation of new anisotropies make $\delta C_l > 0$.\\

These scenarios are addressed in detail in
the next Section, although we stress that this sensitivity to the
relative phase/sign of the fluctuations is {\em intrinsic} to our
method, and applies to any pair of signals.\\

This relative sign dependence leads to the specific (angular) $l$-dependence
of the effects under consideration, and both aspects show up combined
in the final $\delta C_l$'s.\\

\section{Particular Cases and Possible Aplications}

In the context of CMB, the cross term 
$\epsilon \langle a_{l,m} (a^{weak}_{l,m})^* \rangle $ 
discussed above appears due
to different physical processes. 
In what follows, we shall analyse the most relevant in two different 
cosmological scenarios: the $\Lambda$CDM model suggested by WMAP
observations, with cosmological parameters
 $(\Omega_m, \Omega_{\Lambda}, \Omega_{b}, h, n_s)=(0.248, 0.798, 0.044, 0.72, 1.)$, and a critical Einstein-de Sitter Universe with 
$(\Omega_m, \Omega_{\Lambda}, \Omega_{b}, h, n_s)=(0.956, 0, 0.044, 0.72,
1.)$, (hereafter denoted as {\it SCDM} ). The inclusion of SCDM model
responds to the need of understanding the correlations in scenarios with
no ISW effect. \\

The growth of the Large Structure of the Universe is such that 
it is the small overdensities the first ones to become non linear and
form the first haloes,
which, with time, merge to form more massive structures.
In order to see the effect of these haloes on the CMB power spectrum
one must focus on the typical angular distance between sources. 
If sources are distributed uniformly, then one must take into account
only the so-called {\it poissonian} term, but if sources are in some
way clustered, then a {\it correlation} term must be also considered,
\citep{LC93, KK99}. These two contributions conform what we have called
the {\it squared} term, proportional to $\epsilon^2$.

The approach proposed here provides an additional 
way to study the effect of the
halo population on the CMB, consisting in looking at the correlation of their 
spatial distribution with the intrinsic CMB temperature anisotropies; i.e., the
cross ({\it linear in $\epsilon$}) term. This coupling
responds, in most cases ({\em but not all}), to
the correlation of the density fluctuations field with the
gravitational potential fluctuation field 
in a $\Lambda$CDM universe, (ISW effect).
The particular spectral dependence of the cross term compared to the squared
term makes it feasible to distinguish between them, {\em enabling a separate
and independent analysis of the halo population}.\\

We must note that the nature of the correlation is independent of the 
particular physical process, but hinges exclusively on the spatial 
distribution of haloes. 
To model the halo population, we have recurred to the Press-Schechter
formalism, \citep{PS}, which in general provides a good fit to the outcome
of numerical simulations, although small corrections to it have been
suggested, \citep{ShTor, Jenkins}. The latter can be easily 
implemented in our procedure. However, this description of the
halo population must be accompanied by a proper modelling of the
physical environment in the haloes, which condition the physical
phenomena under study, (i.e., the fraction of neutral hydrogen
in 21 cm emission, the cosmological history of the star
formation rate in dust emission, the number density of radio galaxies
versus redshift for radio background studies, etc).
\\

\subsection{Thermal SZ Effect and intrinsic CMB fluctuations}

The tSZ effect arises as a 
consequence of the Doppler change of frequency of CMB photons due
to Thompson scattering on fast moving thermal electrons. 
In this scattering, the transfer of energy
from the electrons to CMB photons translates into a {\em distortion}
of the Black Body spectrum of the CMB radiation. Consequently, 
the tSZ effect introduces frequency dependent temperature anisotropies in the
Cosmic Microwave Background, which, in the non-relativistic limit,
can be written as an integral of
electron pressure along the line of sight,
\begin{equation}
\frac{\delta T}{T_0} = g(\nu ) \int d\eta \; a(\eta)\;\frac{k_B T_e (\eta )}
	{m_e c^2} \sigma_T n_e(\eta ),
\label{eq:tSZ1} 
\end{equation}
with $g(x) \equiv x\coth (x/2) - 4$ and $x\equiv h\nu/k_BT_0$ the
adimensional frequency in terms of the CMB monopole $T_0$. 
For this reason, clusters of galaxies, with their
gravitational wells filled with hot gas 
acting as sources of electron pressure, constitute the
main target of tSZ observations. However, diffuse ionized gas, placed in the
larger scales of superclusters and filaments where still some pressure
support is provided, should also leave an imprint on the CMB spectrum by
means of the tSZ effect. However, this effect is,
for $l < 2000$, remarkably smaller than the intrinsic CMB anisotropies,
and this allows us to apply the formalism outlined above.\\

 Recently there has been active discussions about the origin of 
some {\it excess} power found at $l \;^{>}_{\sim} \;2000$ 
in ground-based CMB 
experiments, \citep{CBI, ACBAR}. Some groups \citep{Bond_tSZ} have argued
that it can be due to tSZ signal coming from unresolved galaxy clusters.
Since the power spectrum is a quantity which, {\it a priori}, does not
retain sign information, methods based on the sign of the skewness 
of the probability distribution function of the signal have been
devoloped in order to discern whether such signal comes from
{\it negative} tSZ clusters or {\it positive} point sources,
\citep{jal03}. \\

In what follows, we show how the frequency dependence of the $\delta C_l$'s
can be of relevance in this problem.
We shall use an approach similar to that
of \citet{Cooray01} to model the temperature fluctuations introduced by the 
population of galaxy clusters. The $k$--mode of the temperature fluctuation
field is given by the following LOS integral: 

\[
\Delta (\vk, \eta_0) =  \;\int d\eta \;g(x) \;
e^{i k\mu[\eta_0 - \eta]}\; \times
\]
\[
\phantom{xxxx}
\biggl[ \int dM \; f(\eta, M) \biggl( \frac{\bar{\rho}}{M} \biggr)^{1/3}
\frac{\sigma_T \bar{n}_e(\eta ) T_e (M, \eta) {\cal D}(M, \eta)}{m_e c^2}
\]
\begin{equation}
\phantom{xxxxxxxxxxxxxxxxxxxxxxxxxx}
  \cdot \; b(M, \eta) \biggr] \;\times \; \delta_{\vk}.
\label{eq:tSZ2}
\end{equation}
$\bar{n}_e(\eta)$ is the background average electron number density at
epoch $\eta$, $T_e (M, \eta)$ is the cluster 
electron temperature given by, e.g., \citet{Eke96}, and $b(M, \eta)$ is
the halo bias factor, \citep{MoW96}. ${\cal D} (M, \eta)$ is density
LOS integral for a $\beta = 2/3$ model (for the case in which
 the line of sight goes through the center of the cluster),
${\cal D}(\eta, M) = 2 r_c (M, \eta) \tan^{-1} (p)$ \citep{Atrio99}, 
with $r_c$ the
cluster core radius the same as used by 
\citet{jal03}, and $p$ is the virial radius to
core radius ratio, which we have taken to be 10.
 The mass integral multiplying $\delta_{\vk}$ represents the pressure 
bias generated at galaxy clusters, and is characterized by the mass
function $f(\eta, M)$ (for which we have used the Press-Schechter (PS)
formalism): 
\begin{equation}
f(\eta, M) = \sqrt{\frac{2}{\pi}} \left| \frac{\partial \sigma }{\partial M} 
	\right|
 \frac{\delta_c}{\sigma^2} \; e^{-\frac{\delta_c^2}{2\sigma^2} },
\label{eq:PSmf}
\end{equation}
where $\delta_c$ is the spherical collapse critical overdensity and 
$\sigma (M, \eta)$ the mass fluctuation field. 

\begin{figure}
\begin{center}
        \epsfxsize=8.5cm \epsfbox{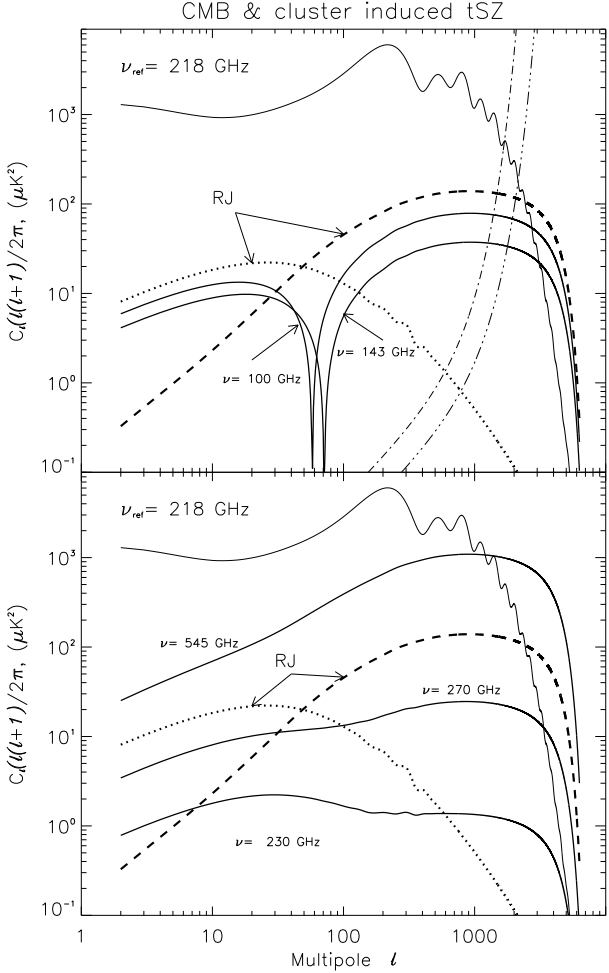}
\caption[fig:dtSZ1]{Sign and frequency dependence of tSZ fluctuations. ({\it In both panels:}) Both dashed and
dotted lines give power spectra in the RJ range, for which the tSZ shows
no frequency dependence. The power spectrum of the 
{\it squared} term associated 
to the tSZ effect due to the
cluster population is given by the dashed line, ($\Lambda$CDM model). 
For the sake of comparison, the 
CMB power spectrum given by the cosmological parameters provided by
WMAP's team is displayed by the upper solid line. 
The dotted line gives the absolute value
amplitude of the cross-correlation term between the intrinsic
CMB and the tSZ signal.
{\it In the  upper panel}, 
the (bottom) solid lines give the actual predicted $|\delta C_l|$'s
obtained after taking a 218 GHz channel as reference, for two close
observing frequencies (100 GHz and 143 GHz). We have assumed that the tSZ
signal cancels exactly in the reference channel, and that the $\delta C_l$'s
are entirely due to tSZ effect. 
Note the change of 
sign of the total power (linear term {\em plus} squared term) 
at $l_{zero}$, below which the linear term dominates, due to the
anticorrelation of tSZ signal and CMB at low frequencies. 
The dot-dashed and the three dot-dashed lines give the nominal amplitude of
the noise residuals for the HFI 143 GHz and 100 GHz channels, respectively.
They are well below the signals we are studying.
{\it In the bottom
panel}, we take 230 GHz, 270 GHz and 545 GHz 
as observing frequencies and 218 GHz as
reference channel, finding no
change of sign for the $\delta C_l$'s, (bottom solid lines).
}
\label{fig:dtSZ1}
\end{center}
\end{figure}

\begin{figure}
\begin{center}
        \epsfxsize=7cm \epsfbox{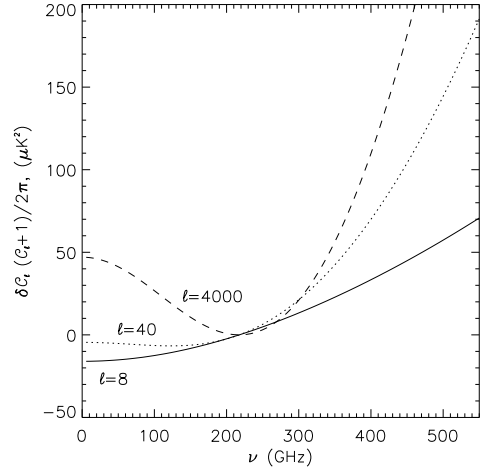}
\caption[fig:dtfreq]{Frequency dependence of the $\delta C_l$'s for
$l=8,40$ and $4000$. For $l=8$, $\delta C_l$ behaves versus frequency
as $g_{tSZ} (x) \equiv x \coth(x/2) -4$, whereas for $l=4000$ the
$\nu$ scaling is proportional to $g_{tSZ}^2$, and never crosses zero.
The $l=40$ is a linear combination of these two extreme cases, weighted
by the relative amplitudes of the linear and squared terms.
}
\label{fig:dtfreq}
\end{center}
\end{figure}

In Figure (\ref{fig:dtSZ1}) we show our results for the $\Lambda$CDM
universe.
The amplitude of the cluster induced tSZ power spectrum (square term 
evaluated at Rayleigh--Jeans (RJ) frequencies) equals
that of the intrinsic 
CMB power spectrum at $l\sim 2000$, and then drops steeply
due to the lack of very high $k$ modes in our integration, (dashed
line). 
Nevertheless, we remark that this 
approach to model the cluster induced signal observes
the effect of the cluster-cluster correlation term, \citep{KK99}, since its
dependence versus $l$ is not $C_l \propto const$ at low multipoles, as it
would be expected for the poissonian term. Provided that, in this
model, cluster induced tSZ temperature fluctuations are determined by the 
matter density fluctuation field, 
its correlation properties are also governed by the matter
power spectrum. Let us also remark that there is no 
flat approximation here, and hence
the predictions should apply to the very large scales. In the small
scales, for which the squared term is dominant, our model is
 comparable to the results of N-body simulations, 
\citep{VolkertSZ, SeljaktSZ, LiPentSZ}, who, compared to each other,
 provide relatively similar predictions.
\\

Our approach aims to describe the interplay between
the linear and the squared terms, together with their combined effect.
However, 
we do not intend to provide accurate predictions for the amplitude of the
tSZ-induced power spectrum: this is an open issue subject to be explored
via hydrodynamical simulations and a better understanding of the distribution
of galaxy clusters with respect to redshift. Progresses at this respect
should leave our qualitative descriptions of the frequency and $l$ dependence
of the tSZ power spectrum untouched.
\\

We can see in figure (\ref{fig:dtSZ1}) that 
the absolute value of the cross term 
evaluated at RJ frequencies (dotted line) shows an
amplitude a factor 5 to 20 higher in the large scales
($l < 20$) than the dashed line (squared term). 
Once the frequency dependence of the cross (linear) and
squared terms is taken into account, we find different patterns for the
$\delta C_l$'s according to the observing frequencies. For $\nu < 218$ GHz,
we see in figure (\ref{fig:dtSZ1}) 
that the $\delta C_l$'s become negative in the low-$l$ range for which the
linear term dominates, and the particular multipole at which $\delta C_l$'s
cross zero (hereafter referred to as $l_{zero}$) 
depends also on the observing frequency. The value of such
multipole for different frequencies in the $\Lambda$CDM model is shown in
figure (\ref{fig:lcross}): it remains roughly constant in the RJ regime,
but approaches higher values as the frequency tends to 218 GHz. This
is due to the fact that the squared term tends to zero much faster than
the linear one when frequencies approach 218GHz. For
 $\nu > 218$ GHz both linear and squared term are positive and hence
$\delta C_l$ does not change sign. {\it Note
that these predictions for the $\delta C_l$'s versus $l$ and frequency
are specific only for the tSZ effect, and should permit to distinguish it from
the contribution of other sources. }\\

\begin{figure}
\begin{center}
        \epsfxsize=7cm \epsfbox{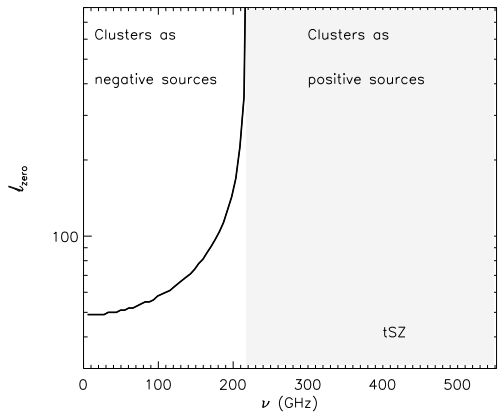}
\caption[fig:lcross]{Multipole at which $\delta C_l$ induced by the tSZ
effect change sigh, for observing frequencies below 218 GHz, for
which clusters can be regarded as {\it negative} point sources,
\citep{korolev86}. Note that
it remains constant for RJ frequencies, but shiftes to larger values as
$\nu$ tends to 218 GHz. Note, however, that at this frequency the tSZ
signal drops to zero.
}
\label{fig:lcross}
\end{center}
\end{figure}

The different dependence versus $\nu$ for different $l$'s is displayed
in Fig.(\ref{fig:dtfreq}): the two extreme cases are given for $\delta C_l$
at $l=8$ and $4000$, whereas the intermediate case corresponds to $l=40$.
The behaviour of the $\delta C_l$'s versus frequency is a consequence
of {\it i)} the independence of the photon spectrum upon redshift and {\it ii)}
the fact that the tSZ surface brightness changes sign at $\nu = 218$ GHz.
\\

After defining the correlation coefficient as ${\cal R}_l \equiv
\langle a_{l,m}^{CMB} a_{l,m}^{tSZ} \rangle /
\sqrt{ C_l^{CMB} C_l^{tSZ}}$, we plot it for both $\Lambda$CDM and SCDM
cosmological models (thick and thin solid lines, respectively),
(figure (\ref{fig:ccoef1})). The ISW is the cause of the coupling
of CMB anisotropies with tSZ signal in the $\Lambda$CDM case. This
causes a cross-correlation with the {\em total} CMB signal of about
a 20\% at $l\sim 10$, which drops at higher multipoles since the ISW
signal decreases rapidly with increasing $l$. For the SCDM model,
we obtain $l_{min} \sim 5$ for $\eta_{tSZ} (z \sim 0.5) \sim 6730$ Mpc
and $\eta_0 \sim 8300$ Mpc;
and $l_{max} \sim 8$, which would explain the low level of correlation
in this case (less than a few percent).\\

\begin{figure}
\begin{center}
        \epsfxsize=7cm \epsfbox{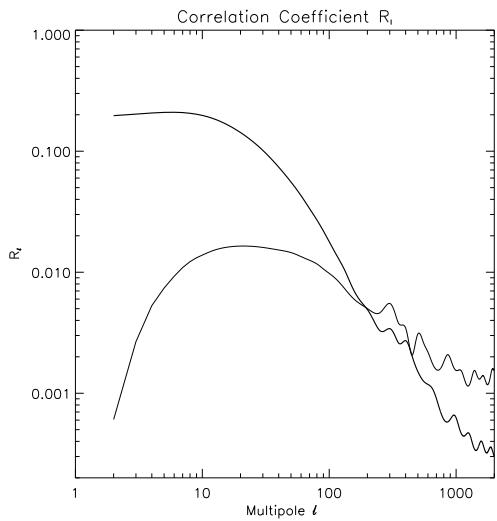}
\caption[fig:ccoef1]{Correlation coefficients ${\cal R}_l \equiv
\langle a_{l,m}^{CMB} a_{l,m}^{tSZ} \rangle /
\sqrt{ C_l^{CMB} C_l^{tSZ}}$ for clusters in the $\Lambda$CDM
 (thick line) and SCDM (thin line) scenarios.
}
\label{fig:ccoef1}
\end{center}
\end{figure}

\subsection{Reionization and Resonant Scattering of CMB Photons on Ions, Atoms and Molecules of Heavy Elements}

Both scattering on free electrons during
reionization and resonant scattering associated to any type of
transition in heavy species contribute with some optical depth  
for the CMB photons. In the first case, the optical depth is
generated by the Thompson scattering occuring between CMB photons and
free electrons, and, hence, is frequency independent. This situation changes
for resonant transitions, provided that
CMB photons scatter the line only if their
frequency is close enough to the resonant frequency. Apart from this
distinction, the effect of both phenomena on the CMB power spectrum is
identical, so we shall restrict our analysis on the case of resonant
scattering, (which, by its spectral peculiarity, 
can be separated from the intrinsic CMB temperature fluctuations).
Hence, we refer to Basu, Hern\'andez--Monteagudo \& Sunyaev 
(2004), (hereafter BHMS) 
where this effect is utilized to discuss constraints on the
abundances of heavy species at redshifts $ 0.1 < z < 30$.

If we denote by $\tau_{rs}$ the {\it homogeneous} (i.e. position independent
\footnote{In the optically thin limit, $\tau_{rs} \ll 1$, one can relax
the approximation on homogeneity by assuming that the scales at which
$\tau_{rs}$ varies are smaller than the scales under study, for which
an {\em average} integrated optical depth is effectively working.})
optical depth associated to resonant 
scattering, we can write that the change induced by the resonant 
transition on the temperature field is given by:
\begin{equation}
T_{rs} = T_{cmb} \; e^{-\tau_{rs}} + T_{gen},
\label{eq:res1}
\end{equation} 
where $T_{rs}$ is the temperature angular fluctuation field at the time of
resonant scattering, $T_{cmb}$ is the intrinsic CMB field generated
at the LSS and $T_{gen}$ are the new temperature
fluctuations generated by the resonantly scattering species. If we now take the
limit $\tau_{rs} \ll 1$, the last equation becomes
\begin{equation}
T_{rs} = (1-\tau_{rs})\; T_{cmb} + \tau_{rs}\; T_{gen}^{lin} + 
	{\cal O} [\tau_{rs}^2],
\label{eq:res2}
\end{equation}
with $T_{gen}^{lin}$ the coefficient of the linear term in the expansion of
$T_{gen}$ in terms of $\tau_{rs}$. In Fourier space, this translates into:
\[
a_{(l,m),\;rs} = (1-\tau_{rs})\; a_{(l,m),\;cmb}\; + \;\tau_{rs}\; 
a_{(l,m)\;gen}^{lin} 
\]
\begin{equation}
\phantom{xxxxxxxxxxxxxxxxxxxxxxxxxxxxxxxxx}
 + \; {\cal O} [\tau_{rs}^2],
\label{eq:res3}
\end{equation}
with $a_{(l,m)}$'s denoting Fourier multipoles.
If we now define $\delta C_l \equiv \langle |a_{(l,m),\;rs} |^2\rangle 
- \langle |a_{(l,m),\;cmb} |^2\rangle$, it is straightforward to
find that
\[
\delta C_l = \tau_{rs} \;2\; \bigl(
\langle {\cal R}e \;\left[ a_{(l,m),\;cmb}\times 
	(a_{(l,m),\;gen}^{lin})^*\right]
 \rangle
\]
\begin{equation}
\phantom{xxxxxxxxxxxxxxxxxxx}
-\langle |a_{(l,m),\;cmb} |^2 \rangle \bigr)
 	\; + \; {\cal O} [\tau_{rs}^2].
\label{eq:res4}
\end{equation}
As shown in detail 
in the Appendix A of BHMS,
the first term accounts for the correlation between fluctuations
generated during recombination and those generated in the epoch
of resonant scattering, whereas the second (autocorrelation) term expresses the
 blurring of the intrinsic anisotropies induced in the LSS
due to the resonant scattering at lower redshift; from now this term
will be referred to as the {\it blurring term}. Note that it is 
merely proportional to the intrinsic CMB power spectrum at the resonant
scattering epoch, and hence, as long as resonant scattering takes place after
recombination, the shape of this blurring term will be identical to the
primordial CMB power spectrum generated at decoupling, and
thus {\em redshift independent}. 
For the reasons outlined at the end of Section 2,
the correlation term is
only of relevance at the very low $l$ range of multipoles, in
which newly generated anisotropies overcome the blurring of
original temperature fluctuations and introduces new anisotropy power,
(see again Appendix A of BHMS).
This occurs for both $\Lambda$CDM and SCDM cosmological models, 
since the Integrated Sachs-Wolfe effect (hereafter ISW) 
has no effect here provided that, in adiabatic $\Lambda$ models,
it becomes important only at very low redshift, during the $\Lambda$ term
dominance, whereas for an Einstein-de Sitter Universe it vanishes in
the linear regime, (e.g., \citet{HuS95}). Recalling that $\eta_{rec} \sim 300$
Mpc, $\eta_{rs} (z = 25) \sim 2811$ Mpc, and that $\eta_0 \simeq 14000$ Mpc,
one finds that $l_{min} \sim 1$ (no drop of the cross correlation
expected at low multipoles) and $l_{max} \sim 30$, at which we would
expect having some decrease in the amplitude and/or change of
sign in the cross correlation. Figure (\ref{fig:dcl1}) shows the
actual computation of the terms in eq. (\ref{eq:res4}): 
all curves have been computed for $\tau_{rs}=10^{-3}$ and 
rescaled to $\tau_{rs}=1$,
so the actual measurement that our method would provide is then
given by the diamonds line {\em times} $\tau_{rs}$, 
({\em which, for small enough
$\tau_{rs}$, is below the cosmic variance limit}). As in BHMS,
the resonant lines
have been modelled by a gaussian centered on the conformal time ($\eta_{rs}$)
corresponding to the redshift considered in each case, and with
a $\sigma$ equal to one percent of $\eta_{rs}$. 
Solid lines
gives the blurring of the original power spectrum, and the dashed
line accounts for the cross-correlation. Note that we are plotting absolute
values, and that only at low multipoles the first term is positive and
greater in amplitude than the blurring term. For higher multipoles,
the correlation term can be neglected and one is left with the simple
autocorrelation term: 
\begin{equation}
\delta C_l \simeq -2 \tau_{rs} \; C_l^{cmb}.
\label{eq:res5}
\end{equation}
This $l$-dependence for the $\delta C_l$'s is generic for any source
of {\it localised} optical depth for the CMB photons. 
This drop in the $\delta C_l$'s (and change of sign at some $l_{zero}$,
see low $l$ range in figure (\ref{fig:dcl1}))
are a direct consequence of the correlation of fluctuations at $\eta_{rec},
\eta_{rs}$, and provide a test for the origin of the $\delta C_l$'s,
just as in the case of the tSZ effect addressed above.

\begin{figure}
\begin{center}
        \epsfxsize=7cm \epsfbox{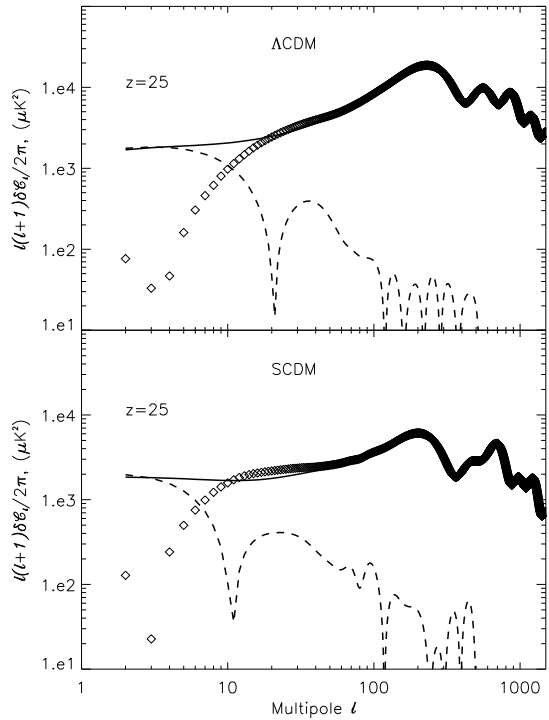}
\caption[fig:dcl1]{
Angular power spectrum arising as a consequence of resonant scattering
on a line placed at the end of the Dark Ages, ($z = 25$). We are
plotting the two terms contributing to the total  $\delta C_l$'s (diamonds):
the absoption term is displayed in thick
solid line, and is merely proportional to the intrinsic CMB power spectrum.
It is negative, and hence the total $\delta C_l$'s cross zero only when
the Doppler-induced generation (dashed line) becomes relevant at low
multipoles. Only absolute values are shown. The correct amplitude 
of this effect is obtained after multiplying these curves by
$\tau_{rs}$.
}
\label{fig:dcl1}
\end{center}
\end{figure}

We remark that these $\delta C_l$'s are measurable only if the CMB 
is being observed at two different frequencies; one
 corresponding to the resonant scattering at $\eta_{rs}$, and another
one in which such resonant scattering can be neglected.  Note that
there is no place for this situation in the case electron scattering
during reionization, since
Thompson scattering on free electrons is frequency independent.
We are implicitely 
assuming that the instrument is sensitive to the amplitudes of
the $\delta C_l$'s: in BHMS we showed that the current detector
technology (present in experiments like WMAP, ACT or Planck) 
should already allow to set strong limits in the abundance
of resonant species during the epoch of reionization.

\subsection{Emission in Fine Structure Lines of C, N, O in Haloes}

BHMS studied the effect of resonant scattering of CMB photons in
fine structure transitions associated to metals and ions. They
found that very overdense regions ($\delta \geq 10^4$) should emit in these
lines via collisional excitations, \citep{spergel_cno, varsh81}. 
The expected
amplitude of this signal is relatively small, while its spectral
dependence is very different from that of the CMB. On the other
hand, it also depends on the star formation history in haloes whose large
scale distribution 
should trace the general density
fluctuation field. For these reasons, one can consider the application of the 
correlation method in this case as well. 
The main difference to the scenario studied by
BHMS is that, in this occasion, the scattering in the lines is almost
negligible, and hence, no blurring of original CMB anisotropies should be
expected. Hence, 
there will be no further suppresion of the CMB power spectrum at 
high multipoles, but only extra power in the large angular scale
range. This is motivation of an upcoming paper where both the linear
and quadratic terms are taken into account.


\subsection{Extragalactic Foregrounds}

In this subsection we address possible
 effects that well-known physical processes
(such as free-free emission, dust emission in the
IGM or inside galaxies and synchrotron emission in
extragalactic radio sources) have on our method. In the case of 
extragalactic foregrounds, it is clear that if they are produced
in haloes, they should trace the overall 
mass distribution in the very large scales, 
{\it just as in our study of tSZ signal induced by
clusters of galaxies}. For this reason, one could think of applying this
method on them, expecting to find a similar
shape for the correlation term at large angular scales as the one 
found for tSZ clusters. This raises the question whether these foregrounds 
could mutually contaminate or bias the correlation estimates.
Since the method proposed here is based on the frequency dependence of the
signal under study, proper frequency coverage should
allow to identify and separate each component as long as spectral
signatures are distinct enough.\\

It is obvious that if the sources of these signals
are located in our Galaxy, one would not expect any type of correlation
between them and the original density perturbation field, {\em leading
to no linear ($\propto \epsilon$) term}.

\section{Discussion and Conclusions}

The amplitude of the cross correlation depends essentially on 
the conformal distance separating the signal sources, rather than
the particular $k$ projection of sources of different origin. The closer the
sources of the signals are, 
the higher the correlation becomes. At this respect, the presence
of a $\Lambda$ term generating an ISW signal is of crucial importance
for those effects generated in our neighborhood, (particularly
the tSZ effect, \citet{Cooray01}). Consistently with the ISW contribution
to the total CMB signal (around 20 $\mu$K with respect the 
total $\sim$ 110 $\mu$K of the CMB), the correlations in
a $\Lambda$CDM universe show typical values of 10-20 \%, with
remarkably lower values in the SCDM scenario. In an Einstein-de Sitter
universe, the correlation drops to a few percent, and the enhancement
of the weak signal is rather far from being relevant.
The situation changes remarkably
in the case of resonant scattering at high redshift. In this situation,
the correlation coefficient is practically unity for the low
multipoles, since, as shown in BHMS, arises as a consequence of the
monopole and Doppler terms of the CMB, and the contribution
 of the ISW component is negligible.\\

Although the galactic contamination is thought to be more important
in the large angular scales where these correlations show up, it is
also expected that space experiments achieve their best sensitivities in 
the big angular scales. 
In the case that the signal is of extragalactic origin, 
the cross term will always show up
together with the squared term, although
both terms have, in general, {\em different} frequency (\ref{eq:deltaC_l_nudp})
and $l$-dependence. This
should also help in distinguishing between them, specially in the case of the
tSZ effect, for which a peculiar pattern of the $\delta C_l$'s versus $l$ and
$\nu$ has been predicted.
 In the high multipole range, frequency dependent
scattering such as resonant scattering introduce a measurable blurring
of original CMB temperature fluctuations generated during recombination.
Since it merely consists in an autocorrelation of CMB anisotropies,
this blurring term has the same l-dependence as the original CMB power
spectrum.\\

The method proposed here can also be applied in the study of the
cross correlation of CMB temperature fluctuations with the radio
background. In the low frequency range, new instruments like the 
Low Frequency Array (LOFAR) or the  
Square Kilometer Array (SKA) will measure the radio background. This is 
mainly due to radio galaxies present in the redshift range $z\in [0,4]$,
and its fluctuations are expected to be of much higher amplitude than 
those of the CMB.
However, due to the fact that the radio background is generated by
radio galaxies tracing the universal density fluctuation field, one
can think of applying this method in order to enhance the CMB
component at these frequencies. When doing this, one must keep in 
mind that there is emission at 21 cm coming from neutral hydrogen
during the Dark Ages, ($z \sim (30,100)$, \citet{Madau21}) 
 which should fall in this
frequency range and which is showing also some degree of 
correlation with the CMB. However, according to the arguments given
in Section 2, most of the correlation will be due to the coupling of the
ISW effect with the radio galaxy distribution at low and moderate
redshifts. \\

Similar arguments can be applied when studying the 857 GHz band of
Planck's HFI, since we can expect that this method should be able to
unveil the distribution of extragalactic dust and its imprint on the
CMB. In other words, by means of the ISW the CMB has become a tool
which permits performing independent tests at different frequencies on
the large scale distribution of matter. The main two caveats to have
present are the possibility of having some signal generated during
reionization, at very high redshift, which could be introducing
some extra correlation, and the presence of galactic foregrounds,
whose residuals might invalidate these analyses in the very
low multipoles. \\ 

In this paper, we have addressed the issue of correlated signals in
the context of CMB. We have shown that, in the case in which two signals
have different spectral dependence, the presence of correlations between
both can be used in order to enhance the weak signal with respect the
dominant one. Assuming that the correlation between signals is caused by the
the cosmological density perturbation field, we have found at 
which angular range such correlation might be relevant. This depends
essentially on two different scales: the distance separating the events 
generating the signals under consideration, and their distance to the
observer. In a $\Lambda$CDM universe, 
these cross terms dominate at the large angular scales, and 
hence characterize our predictions of the power spectra associated to the
weak signals in the low multipole range.\\

\section*{Acknowledgments} 

C.H.M acknowledges the financial support provided through the European
Community's Human Potential Programme under contract HPRN-CT-2002-00124,
CMBNET. The authors acknowledge F.Atrio--Barandela and K.Basu for 
stimulating discussions, and J.A.Rubi\~no--Mart\1n for reading the manuscript.
The authors also acknowledge discussions with L.Page, J.L.Puget and 
A.Readhead, which increased authors' confidence
 that future experiments will permit to observe
the effects discussed in this work.

\label{lastpage}

\end{document}